\documentclass[12pt]{article}
\usepackage{amsfonts,amssymb,amsmath}
\usepackage[dvips]{epsfig}
\begin{document}
\title{Landau Levels as a Limiting Case
of a Model with the Morse-Like Magnetic Field  }
\author{ H. Fakhri \\ Department of Theoretical Physics and
Astrophysics, Physics Faculty,\\ University of Tabriz, Tabriz 51666-16471, Iran
\\ e-mail: hfakhri@tabrizu.ac.ir
\\[2ex] B. Mojaveri\\Department of Physics, Faculty of Science, Razi University,
Kermanshah, Iran\\and
\\  Department of Physics, Azarbaijan University of Tarbiat Moallem, \\ Tabriz 53741-161, Iran
\\ e-mail: bmojaveri@raziu.ac.ir \\
[2ex] M.A. Gomshi Nobary\\
Department of Physics, Faculty of Science, Razi University,
Kermanshah, Iran\\ e-mail: mnobary@razi.ac.ir } \maketitle
\begin{abstract}
We consider the quantum mechanics of an electron trapped on an infinite band along the $x$-axis in
the presence of the Morse-like perpendicular magnetic field $\vec{B}=-B_{0}e^{-\frac{2\pi}{a_{0}}x}\hat{k}$
with $B_{0}>0$ as a constant strength and $a_{0}$ as the width of the band.
It is shown that the square integrable pure states realize representations of $su(1,1)$ algebra via
the quantum number corresponding to the linear momentum in the $y$-direction.
The energy of the states increases by decreasing the width $a_{0}$ while it is not changed by $B_{0}$.
It is quadratic in terms of two quantum numbers, and the linear spectrum of the Landau levels
is obtained as a limiting case of $a_{0}\rightarrow\infty$.
All of the lowest states of the $su(1,1)$ representations minimize uncertainty relation and
the minimizing of their second and third states is transformed to that of the Landau levels in
the limit  $a_{0}\rightarrow\infty$. The compact forms of the Barut-Girardello coherent states corresponding to
$l$-representation of $su(1,1)$ algebra and their positive definite measures on the
complex plane are also calculated.

\end{abstract}

\section{Introduction}
The physics of charged particles in a magnetic field
has been one of the important problems in various fields such as
condensed matter physics, quantum optics etc. The discrete
energy values corresponding to motion of a charged particle on the
infinite flat surface in the presence of a uniform magnetic field
perpendicular to this plane are called Landau levels
\cite{Landau,Ferreyra,Rohringer,Fakhri1,Mouayn,Yang,Bracken,Twareque}. The energy levels are linearly dependent
on the mode and consequently the successive energy gaps have the
same values. The diamagnetic property of a metal is described by
such considerations. Also, the Landau problem for
various manifolds such as the sphere $S^2$, hyperbolic plane $H^2$
and four-dimensional sphere $S^4$ with symmetry groups $SO(3)$,
$SO(2,1)$ and $SO(5)$ have been considered
\cite{Wu,Dunne,Zhang}.

From the classical viewpoint, the magnetic field creates a
transverse current along perpendicular to both the direction of
motion of particle and the direction of the magnetic field.
Therefore, the Landau problem can be counted as a cornerstone of
quantum Hall effect. In the last three decades, many efforts for
describing the spectral properties of the quantum Hall effect have
been carried out
\cite{Laughlin,Wybourne,Chakraborty,Bagarello1,Richter,Bagarello2,Murthy,Karabali,Raya,Hasebe}.
The quantum Hall effect as a universal phenomenon is observed on
any two-dimensional surface with a charged particle moving in the
presence of a strong perpendicular uniform magnetic field. The
quantum Hall conductivity is quantized by the ratio between the
number of electrons and the degeneracy of the Landau levels, the
so-called filling factor $\nu$. The strength of the magnetic
field is in the sense of large energy gap between adjacent Landau
levels, and it leads to the fact that $\nu$ becomes integer valued
and the first $\nu$ Landau levels are completely filled. In this
case, the strong magnetic field does not allow the electrons to
interact with each other. In the case of the weak magnetic field,
$\nu$ is non-integer and the Landau levels are partially filled.

Also, the motion of a charged particle on a two-dimensional surface with the different boundary conditions
and interacting with a magnetic field perpendicular to it is of interesting problems of quantum mechanics.
The Iwatsuka model with the continuous spectrum and the finite number of open spectral gaps is one of
the models in this area  \cite{Iwatsuka,Mantoiu,Exner1}.
The other models that can be mentioned are: the interaction of a
spinless charged two-dimensional particle with a perpendicular homogeneous magnetic field
and the different potentials, e.g., a periodic array of point obstacles,
a potential wall which is transported along a closed loop in the plane, and
a periodic lattice of point perturbations \cite{Exner2,Exner3,Exner4}.
Such models can be interpreted as the modified or perturbed versions of the Landau levels.
In this paper, the surface of the Landau problem is considered as
an infinite flat band with a finite width, and the magnetic field
is chosen as an exponential function of the lengthwise coordinate
of the band. We find exact expressions for the pure states and
values of the energy and show that the infinite degeneracy of the
Landau levels is broken here. It is shown that the successive
energy gaps have not the same values since instead of a linear
spectrum, we obtain a quadratic spectrum in terms of two quantum
numbers and thus, the quantum Hall conductivity can be considered
with more complex properties.

\section{The model}
Consider a spinless electron with charge $e<0$ and an effective mass $\mu$ moving on an
infinite flat band in the presence of a Morse-like perpendicular magnetic field directed in the negative
$z$-direction:
\begin{eqnarray}\label{1}
\vec{B}=-B_{0}e^{-\frac{2\pi}{a_{0}}x}\hat{k},\hspace{30mm}
-\frac{a_{0}}{2}\leq y\leq \frac{a_{0}}{2},
\end{eqnarray}
with $B_{0}>0$ as a constant magnetic field.
It is translationally invariant in the $y$-direction and is assumed that
the electron is unable to enter the barriers and it is compelled to move on the
infinite band surrounded by impenetrable barriers. Therefore, the
wavefunction is zero everywhere in the barriers and this problem
concentrates on an electron trapped on the infinite band along the
$x$-axis. The strength of magnetic
field decreases exponentially with increasing $x$ and remains
unchanged when the coordinate $y$ changes.
The $y$-independency of the Hamiltonian makes it to commute with the momentum operator
$\hat{p}_{y}$, thus for such a system the wavefunction must be
continuous at the points opposite each other on the boundaries:
$\psi(x,y)|_{y=\frac{a_{0}}{2}}=\psi(x,y)|_{y=-\frac{a_{0}}{2}}$.
This is equivalent to
the periodic boundary condition in the $y$-direction
and is in agreement with the fact that the magnetic field is independent of $y$.
Applying the boundary condition on the $y$-axis,
it becomes obvious that the $y$-dependence of $\psi(x,y)$ is as $e^{-i\frac{2\pi}{a_{0}}n y}$
with $n$ as an integer number.

We show that the condition of square integrability on the infinite
band in the presence of magnetic field $\vec{B}$ requires the pure
states to be labeled with another integer quantum number, indicated by
$l$, with the limitation $0\leq l \leq n-1$. $su(1,1)$ Lie algebra
is represented by the pure states via the quantum number $n$,
despite the fact that this problem does not contain a dynamical
symmetry group $SU(1,1)$. However, the commutativity of the
momentum operator $\hat{p}_{y}$ and  the Casimir operator
$su(1,1)$ with the Hamiltonian are known as responsible for the
generation of quantization of $n$ and $l$.
Our considerations show
that the maximum degeneracy possible for the energy levels is
two-fold and the energies increase by decreasing the width
$a_{0}$. As seen, the magnetic field $\vec{B}$ tends to the
constant $-B_{0}\hat{k}$ for $a_{0}\rightarrow\infty$, and so
we will find that the linear spectrum of Landau levels can be
obtained as a limiting case of the model.
Furthermore, we will show that the uncertainty relation not only is minimized on the lowest bases
of the representations of the $su(1,1)$ algebra but also the deviations of the
uncertainty for the second and third bases are transformed to the known deviations of the uncertainty on Landau levels,
in the limit $a_{0}\rightarrow\infty$.
It is also shown that the $l$-representation of $su(1,1)$ generates the coherency of
$su(1,1)$-Barut-Girardello type with a positive definite measure
to satisfy resolution of unity.
\section{Schr\"odinger wavefunctions}
Using CGS units, the Morse-like magnetic field (\ref{1})
can be obtained from the vector potential
\begin{eqnarray}\label{2}
A_x=\frac{i\pi \hbar c}{e\, a_0}-\frac{iB_0\, a_0}{2\pi}\,e^{-\frac{2\pi}{a_0} x},\hspace{15mm}
A_y=\frac{B_0\,a_0}{2\pi}e^{-\frac{2\pi }{a_0} x},\hspace{15mm}A_z=0,
\end{eqnarray}
in which $c$ is the velocity of electromagnetic waves in the vacuum.
Note that $A_x$ can be made equal to zero in the Landau gauge.
For an electron of effective mass $\mu$ moving on the infinite
flat band in the presence of the vector potential (\ref{2}),
the time-independent Schr\"odinger wave equation
\begin{eqnarray}\label{3}
\frac{1}{2\mu}\left[\left(\hat{p}_{x}-\frac{e}{c}A_x\right)^2+
\left(\hat{p}_{y}-\frac{e}{c}A_y\right)^2\right]\psi=E\psi,
\end{eqnarray}
can be written in terms of the variable $\xi=e^{\frac{2\pi}{a_0} x}$ ($0<\xi<\infty$) as
\begin{eqnarray}\label{4}
\left[-\xi^2\frac{\partial^2}{\partial
\xi^2}+\left(\frac{e B_0 a_0^2}{2\pi^2\hbar\, c}-2\xi\right)\frac{\partial}{\partial \xi}
-\frac{a_{0}^{2}}{4\pi^2}\frac{\partial^2}{\partial y^2}+i\frac{e B_0 a_{0}^{3}}{4 \pi^{3} \hbar\, c}
\frac{1}{\xi}\frac{\partial}{\partial y}-\frac{1}{4}\right]\psi=
\frac{2\mu\, a_{0}^{2}}{4\pi^2\hbar^2}E \psi.
\end{eqnarray}
The boundary condition in the $y$-direction requires that the wavefunction $\psi$
be separated into a product of two functions, one that depends only on $y$ and another only on $\xi$:
$\psi=e^{-i\frac{2\pi}{a_0}n y}\psi(\xi)$. Hence, $\psi(\xi)$
will be the solution of the following differential equation
\begin{eqnarray}\label{5}
\xi^2\frac{d^2\psi(\xi)}{d\xi^2}+\left(2 \xi-\frac{e B_0\, a_{0}^{2}}{2\pi^2 \hbar\, c}\right)\frac{d\psi(\xi)}{d\xi}
-\left(n^2-\frac{1}{4}+\frac{e B_0\, a_{0}^{2}}{2\pi^2\hbar\, c}\frac{n}{\xi}-
\frac{2\mu\, a_{0}^{2}}{4\pi^2\hbar^2}E\right)\psi(\xi)=0.
\end{eqnarray}
The square integrable solutions can be obtained by comparing the last equation with
the associated Bessel differential equation of Ref. \cite{Fakhri2}:
$\psi(\xi)=B_{l,n}^{(0,\frac{-e B_{0}\,a_{0}^{2}}{2\pi^2\hbar\, c})}(\xi)$ in which
the associated Bessel functions are given by
\begin{eqnarray}\label{6}
&&\hspace{-10mm}B_{l,n}^{(0,\beta)}(\xi):=
\beta^l\sqrt{\frac{\Gamma(n-l)}{\Gamma(n+l+1)}}\,\xi^{-l-1}L_{n-l-1}^{(2l+1)}\left(\frac{\beta}{\xi}\right)
\nonumber\\
&&\hspace{6mm}=\frac{\beta^{-l-1}(-1)^{n-l-1}}{\sqrt{\Gamma(n+l+1)\Gamma(n-l)}}\,\xi^{n}
e^{\frac{\beta}{\xi}}\left(\frac{d}{d\xi}\right)^{n+l}\left(\xi^{2l}e^{\frac{-\beta}{\xi}}\right),
\end{eqnarray}
with $\beta$ as an arbitrary positive real number. Therefore, the pure states
whose probability density is independent of $y$,
are calculated as\footnote{Instead of (2) we can also use a more general expression for vector potential,
namely:
$$
A_x=\frac{i\pi \hbar c}{e\, a_0}(1-q)-\frac{iB_0\, a_0}{2\pi}\,e^{-\frac{2\pi}{a_0} x},\hspace{15mm}
A_y=\frac{\pi \hbar c}{e\, a_0}q+\frac{B_0\,a_0}{2\pi}e^{-\frac{2\pi }{a_0} x},\hspace{15mm}A_z=0.
$$
In this case, the wavefunctions $\psi_{l,n}(x,y)$ are expressed in terms of the associated Bessel functions
$B_{l,n}^{(q,\frac{-e B_{0}\,a_{0}^{2}}{2\pi^2\hbar\, c})}\left(e^{\frac{2\pi}{a_0}x}\right)$.
Since $q$ is not determined by the boundary condition of the model,
it may be regarded as a gauge degree of freedom in a sense.
Thus, we have fixed the gauge by setting $q=0$.}
\begin{eqnarray}\label{7}
\left|l,n\right\rangle:=
\psi_{l,n}(x,y)=\sqrt{\frac{-e B_0}{\pi\hbar\, c}\left(2l+1\right)}\,
e^{-i\frac{2\pi}{a_0}n y}
B_{l,n}^{(0,\frac{-e B_{0}\,a_{0}^{2}}{2\pi^2\hbar\, c})}\left(e^{\frac{2\pi}{a_0}x}\right).
\end{eqnarray}
The square integrability conditions $n\geq 1$ and $0\leq l\leq n-1$
for the bound-state wavefunctions $\psi_{l,n}(x,y)$ can be deduced from Ref. \cite{Fakhri2}.
They form an orthonormal set with respect to both integer indices $l$ and $n$, that is,
(bar is for the complex conjugation)
\begin{eqnarray}\label{8}
\left\langle l,n|l^{\prime},n^{\prime}\right\rangle:=
\int^{\frac{a_0}{2}}_{-\frac{a_0}{2}}\int^{\infty}_{-\infty}
\overline{\psi_{l,n}(x,y)}\,\psi_{l^{\prime},n^{\prime}}(x,y)\,
e^{\frac{2\pi}{a_0}x}e^{\frac{e B_{0} a_{0}^{2}}{2\pi^2\hbar\, c}
\,e^{-\frac{2\pi}{a_0}x}}dydx=\delta_{l\,l^{\prime}} \delta_{n\,n^{\prime}}.
\end{eqnarray}
Let us here denote the Hilbert space corresponding to all square integrable pure states
with ${\cal H}=\mbox{span}\{\left|l,n\right\rangle,\,\,\,n\geq 1,\,\,0\leq l\leq n-1\}$.
Defining the nonnegative integer number $N:=n-l-1$, ${\cal H}$
can be split into the infinite direct sums of infinite-dimensional Hilbert subspaces in two different
ways: ${\cal H}=\oplus_{N=0}^{\infty}{\cal H}_{N}$ with
${\cal H}_{N}=\mbox{span}\{\left|l,l+N+1\right\rangle\}_{l=0}^{\infty}$ and
${\cal H}=\oplus_{l=0}^{\infty}{\cal H}^{l}$ with
${\cal H}^{l}=\mbox{span}\{\left|l,l+N+1\right\rangle\}_{N=0}^{\infty}$.
In Figure 1, we have schematically shown the bases of the Hilbert space ${\cal H}$ as the points
$(l,n)$ on a flat plane whose horizontal and vertical axes are labeled with $l$ and $n$, respectively.
The Hilbert subspace ${\cal H}_{N}$ involves all pure states situated on
$n=l+N+1$ oblique line. Also, according to our considerations, all pure states settled on the
$l$-th vertical line are denoted by the Hilbert subspace ${\cal H}^{l}$.\\\\
\begin{figure}
\centering
\includegraphics[width=445 pt]{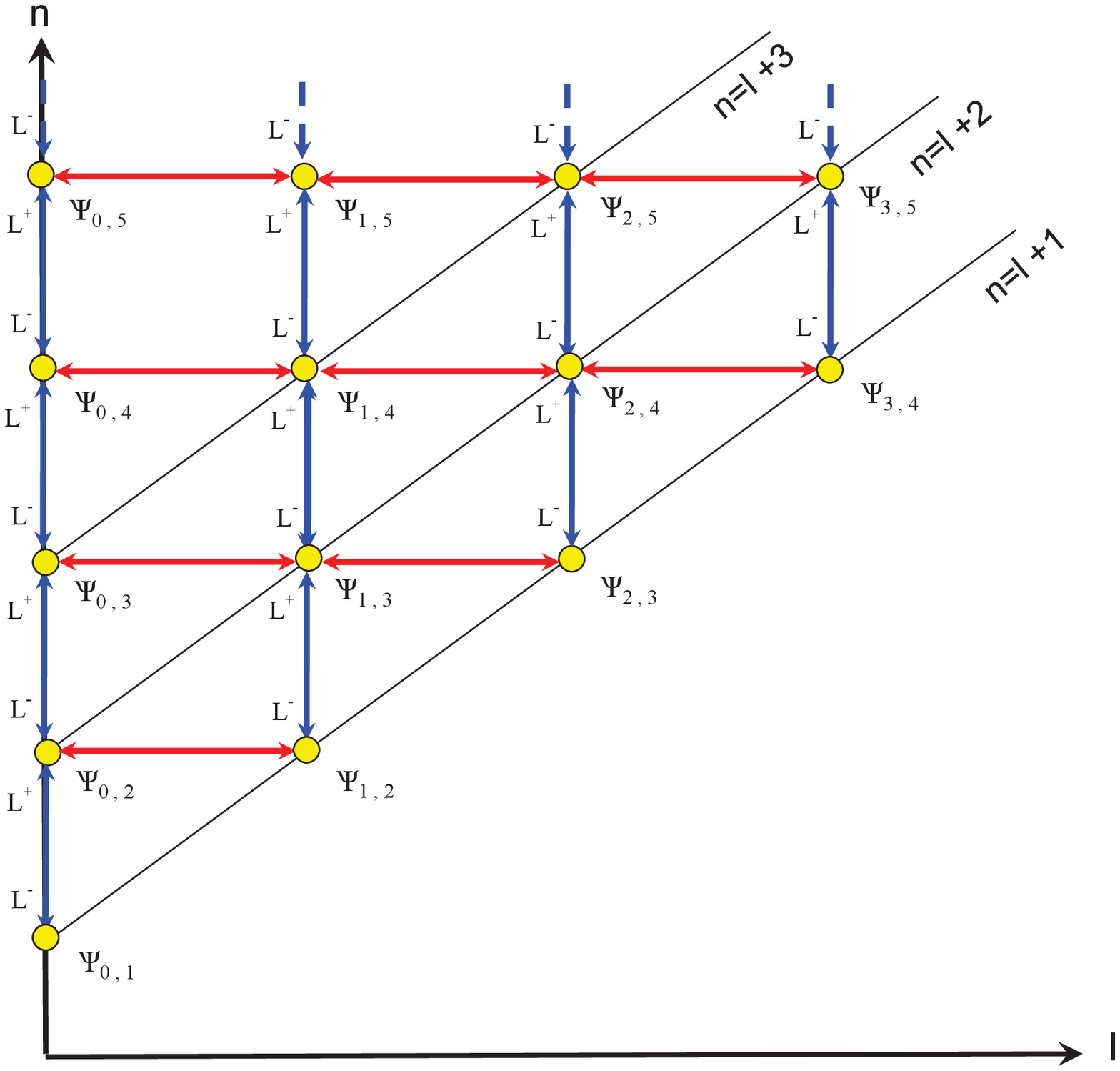}
\caption{} \label{Fig1}\itemize{}
\item Figure. 1. {The plane of displacements of allowed quantum states of the model
with Morse-like magnetic field. The quantum states placed on the vertical lines
constitute the infinite-dimensional irreducible unitary representation spaces of the $su(1,1)$ algebra.
In the limit $l\rightarrow\infty$, the quantum  states settled on the oblique
line $n=l+N+1$ collapse to the Landau level with the energy
$E_{N}=\frac{\hbar \left|e\right| B_{0}}{\mu c}(N+\frac{1}{2})$.}
\end{figure}

From our comparison with the Ref. \cite{Fakhri2} we also find that the allowed energies of
the electron are quantized as a positive quadratic function of both quantum numbers $l$ and $n$:
\begin{eqnarray}\label{9}
E_{l,n}=\frac{2\pi^2\hbar^2}{\mu\,a_0^2}
\left(n-l-\frac{1}{2}\right)\left(n+l+\frac{1}{2}\right).
\end{eqnarray}
Therefore, the energy values increase with decreasing the width $a_{0}$ without
increasing the strength of magnetic field.
In addition, according to our considerations, there is no degeneracy in the case where
$\left(2n-2l-1\right)\left(2n+2l+1\right)$ is a prime number while it is two-fold in other cases.
The ground state $\left|0,1\right\rangle$ is the state with the highest stability and lowest energy:
$E_{0,1}=\frac{3\pi^2\hbar^2}{2\mu\,a_0^2}$.

\section{$su(1,1)$ realization and Barut-Girardello coherent states}
From equation (12b) of Ref. \cite{Fakhri2} it becomes obvious that the $su(1,1)$ Lie algebra
\begin{eqnarray}\label{10}
\left[L_+,L_-\right]=-2L_3,\quad\quad\left[L_3,L_{\pm}\right]=\pm L_{\pm},
\end{eqnarray}
is represented by the quantum states $\left|l,n\right\rangle$:
\begin{eqnarray}
&&L_{+}\left|l,n-1\right\rangle=\sqrt{\left(n+l\right)\left(n-l-1\right)}\left|l,n\right\rangle,\nonumber \\
&&L_{-}\left|l,n\right\rangle=\sqrt{\left(n+l\right)\left(n-l-1\right)}\left|l,n-1\right\rangle,\nonumber\\\label{11}
&&L_{3}\left|l,n\right\rangle=n\left|l,n\right\rangle,
\end{eqnarray}
where the differential explicit forms of the operators are calculated as
\begin{eqnarray}
&&L_+=\frac{a_0}{2\pi}e^{-i\frac{2\pi}{a_0}y}\left(-\frac{\partial}{\partial
x}+i\frac{\partial}{\partial y}+\frac{e B_0\, a_0}{\pi\hbar\, c}\,e^{-\frac{2\pi}{a_0}x}\right),\nonumber\\
&&L_-=\frac{a_0}{2\pi}e^{i\frac{2\pi}{a_0}y}\left(\frac{\partial}{\partial
x}+i\frac{\partial}{\partial y}\right),\nonumber \\\label{12}
&&L_3=i\frac{a_0}{2\pi}\frac{\partial}{\partial y}=:\frac{-a_{0}}{2\pi\hbar}\hat{p}_{y}.
\end{eqnarray}
Note that $L_{3}$ is a self-adjoint operator, and the two operators $L_{+}$ and $L_{-}$ are
Hermitian conjugate of each other with respect to the inner
product (\ref{8}). Therefore, the relations (\ref{11}) are an $l$-integer unitary irreducible
representation of the $su(1,1)$ algebra given in Eqs. (\ref{10}).
For a given $l$, the operators $L_+$ and $L_-$ increase and decrease
energy of the pure states, respectively.

By rewriting the Schr\"odinger equation (\ref{3}) in terms of the $su(1,1)$ generators,
\begin{eqnarray}\label{13}
H=\frac{2\pi^2 \hbar^2}{\mu\, a_0^2}\left(L_+L_-+L_3-\frac{1}{4}\right),\quad\quad\quad\quad
H\left|l,n\right\rangle=E_{l,n}\left|l,n\right\rangle,
\end{eqnarray}
it becomes obvious that the Hermitian Hamiltonian has no dynamical symmetry group $SU(1,1)$, i.e., we have
$[H,L_+]\neq 0$ and $[H,L_-]\neq 0$. However,  the irreducible
$l$-representations are determined by the eigenvalues of the Casimir operator
$C=L_{+}L_{-}-L_3^2+L_3$:
\begin{eqnarray}\label{14}
C\left|l,n\right\rangle=-l(l+1)\left|l,n\right\rangle.
\end{eqnarray}
Therefore, the pure states placed on the $l$-th vertical line of Figure 1,
i.e. the orthonormal bases of ${\cal H}^{l}$,
constitute an infinite-dimensional irreducible $l$-representation of the $su(1,1)$ algebra.
The commutative relations $[H,L_3]=0$ and $[H,C]=0$ are
responsible for the quantization of the energy levels via the
quantum numbers $n$ and $l$, respectively. A basic characteristic
of the system is the following: not only ground state
$\left|0,1\right\rangle$ but also all states
$\left|l,n\right\rangle$ with $n=l+1$, are  annihilated as
$L_{-}\left|l,l+1\right\rangle=0$. Therefore, pure states
belonging to ${\cal H}_{0}$ are the lowest states of
$l$-representations of $su(1,1)$ algebra.

Using the Barut-Girardello eigenvalue equation for the lowering operator, i.e., $L_{-}|Z\rangle_l=Z|Z\rangle_l$,
the normalized coherent states are calculated as
\begin{eqnarray}
&&\hspace{-15mm}|Z\rangle_l=\frac{\left|Z\right|^{l+\frac{1}{2}}}{\sqrt{I_{2l+1}\left(2\left|Z\right|\right)}}
\sum_{N=0}^{\infty}\frac{Z^N}{\sqrt{\Gamma\left(N+1\right)\Gamma\left(2l+N+2\right)}}
\left|l,l+N+1\right\rangle\nonumber \\ \label{15}
&&\hspace{-12mm}=\frac{\sqrt{2\pi(2l+1)}}{a_0}\,e^{-\frac{\pi}{a_0}\left(x+iy\right)
+Ze^{-i\frac{2\pi}{a_0}\,y}}\left(\frac{\left|Z\right|}{Z}\right)^{l+\frac{1}{2}}\frac{J_{2l+1}
\left(2\sqrt{\frac{-B_0 e a_0^2}{2\pi^2 \hbar\,
c}Z\,e^{-\frac{2\pi}{a_0}(x+iy)}}\right)}{\sqrt{I_{2l+1}\left(2\left|Z\right|\right)}},
\end{eqnarray}
in which the J-Bessel functions and the modified Bessel functions of the first kind are as
\begin{eqnarray}\label{16}
J_{2l+1}(u)=\sum_{N=0}^{\infty}\frac{(-1)^N(\frac{u}{2})^{2l+2N+1}}{\Gamma\left(N+1\right)\Gamma\left(2l+N+2\right)},
\hspace{2mm}
I_{2l+1}(u)=\sum_{N=0}^{\infty}\frac{(\frac{u}{2})^{2l+2N+1}}{\Gamma\left(N+1\right)\Gamma\left(2l+N+2\right)}.
\end{eqnarray}
$Z$ is an arbitrary complex variable with the polar form $Z=re^{i\varphi}$ so that $0\leq r<\infty$ and
$0\leq\varphi< 2\pi$.   In addition, the following relations have been used to derive (\ref{15})
\cite{Fakhri2,Gradshteyn}:
\begin{eqnarray}
&&\hspace{-7mm}B_{l,\,n+l+1}^{\left(\frac{-eB_0a_0^2}{2\pi^2\hbar\, c}\right)}\left(e^{\frac{2\pi}{a_0}x}\right)=
\left(\frac{-eB_0a_0^2}{2\pi^2\hbar\, c}\right)^l\sqrt{\frac{\Gamma\left(n+1\right)}{\Gamma\left(2l+n+2\right)}}
\,e^{-\frac{2\pi}{a_0}\left(l+1\right)x}L_n^{\left(2l+1\right)}
\left(\frac{-eB_0a_0^2}{2\pi^2\hbar\,c}e^{-\frac{2\pi}{a_0}x}\right),\nonumber
\\ \label{17} \\ \label{18}
&&\hspace{-7mm}
J_\alpha\left(2\sqrt{uv}\right)e^v\left(uv\right)^{-\frac{\alpha}{2}}=
\sum_{n=0}^{\infty}\frac{v^n}{\Gamma(n+\alpha+1)}L_n^{(\alpha)}(u),\quad\quad\quad\quad \alpha>-1.
\end{eqnarray}
Let us denote the identity operator on subspace ${\cal H}^{l}$ by $I_{l}$.
In order to realize the property of resolution of the identity
$\int d\mu_{_{l}}(Z)|Z\rangle_{l\,\,l}\langle Z|=I_{l}$, we introduce the positive definite
measure $d\mu_{_{l}}(Z)=\frac{2}{\pi}I_{2l+1}(2r)K_{-2l-1}(2r)rdrd\varphi$ with
$K_{-2l-1}(2r)$ as the modified Bessel function of the second kind:
\begin{eqnarray}\label{19}
K_{-2l-1}(2r)=\frac{\Gamma\left(-2l-\frac{1}{2}\right)(4r)^{-2l-1}}{\sqrt{\pi}}\int^{\infty}_{0}\frac{\cos t
}{\left(t^2+4r^2\right)^{-2l-\frac{1}{2}}}\,dt,
\end{eqnarray}\label{20}
which satisfies the following integral relation:
\begin{eqnarray}
\int^{\infty}_{0}r^{2l+2n+2}K_{-2l-1}(2r) dr=\frac{1}{4}\Gamma(n+1)\Gamma(2l+n+2).
\end{eqnarray}

\section{Concluding remarks: Landau levels as a limiting case of the model}
\begin{itemize}

\item  In the limit $a_{0}\rightarrow\infty$, the infinite flat band and the
Morse-like magnetic field are transformed to the infinite flat surface and the uniform
magnetic field perpendicular to this plane, respectively. Therefore, it is expected that
the Landau levels  are obtained as a special case of (\ref{9}). Let $l$ and $n$
tend to the infinity, i.e. $l$ and $n\rightarrow\infty$,
such that $l-n=-N-1$ is the constant.
It is easy to show that the linear relation of Landau levels with quantum number, i.e.
$E_{l,n}\rightarrow E_{N}=\frac{\hbar \left|e\right| B_{0}}{\mu c}(N+\frac{1}{2})$,
is obtained from the limit $a_{0}=2\pi\sqrt{\frac{\hbar\,c\,l}{\left|e\right| B_{0}}}\rightarrow\infty$.
So, the quantum numbers $l$ and $n$ are replaced by  one discrete quantum number $N$
due to the quantization of Landau levels via the Hermite polynomials (see equations (\ref{24})).
This implies that for any Landau level $E_{N}$ in the asymmetric gauge there exists
an infinite-dimensional Hilbert subspace
${\cal H}_{N}$ of discrete pure states of the model which collapses to this Landau level
in the limit $l\rightarrow\infty$.

\item  The strongest criterion to describe difference between the
behavior of quantum and classical mechanics is the
Schr\"odinger-Robertson uncertainty relation \cite{Spiridonov}:
$\Delta:=\sigma_{xx}\sigma_{pp}-\sigma_{xp}^2\geq\frac{\hbar^2}{4}$,
where $\sigma_{ab}=\frac{1}{2}\left\langle
ab+ba\right\rangle-\left\langle a\right\rangle\left\langle
b\right\rangle$, and the angular brackets denote the expectation
value over an arbitrary normalized state. If we compute the
expectation values on the lowest states
$\left|l,l+1\right\rangle$, then we obtain the following results
\begin{eqnarray}
&&\hspace{-10mm}\left\langle x\right\rangle_{l,l+1}=\frac{-a_{0}}{2\pi}\left(\Psi(2l+1)-
\ln\left(\frac{-eB_{0}a_{0}^2}{2\pi^2\hbar c}\right)\right),
\hspace{13mm}\left\langle p_{x}\right\rangle_{l,l+1}=i\hbar\frac{2\pi}{a_{0}}(l+1),\nonumber\\
&&\hspace{-10mm}\left\langle x^2\right\rangle_{l,l+1}=
\left\langle x\right\rangle_{l,l+1}^2+\frac{a_{0}^2}{4\pi^2}\zeta(2,2l+1),
\hspace{30mm}\left\langle p_{x}^2\right\rangle_{l,l+1}=\left\langle p_{x}\right\rangle_{l,l+1}^2,
\nonumber\\ \label{21}
&&\hspace{-10mm}\left\langle xp_{x}\right\rangle_{l,l+1}=-i\hbar(l+1)\left(\Psi(2l+1)-
\ln\left(\frac{-eB_{0}a_{0}^2}{2\pi^2\hbar c}\right)\right),
\end{eqnarray}
in which $\Psi(s)=\frac{d\ln\Gamma(s)}{ds}$ is the logarithmic
derivative of the Gamma function and
$\zeta(2,s)=\frac{d^2\zeta(s)}{ds^2}$ is the second order
derivative of the Riemann zeta function. We have used in (\ref{21}) the
following integral relations \cite{Gradshteyn} ($Re\mu, Re\nu>0$)
\begin{eqnarray}\label{22}
&&\hspace{-15mm}\int_{0}^{\infty}s^{\nu-1}e^{-\mu s}\left(\ln s\right)^j\, ds=\frac{\Gamma(\nu)}{\mu^{\nu}}
\left(\left(\Psi(\nu)-\ln\mu\right)^{1+\delta_{j\,2}}+\zeta(2,\nu)\delta_{j\,2}\right),
\hspace{4mm}j=1,2.
\end{eqnarray}
The relations (\ref{21}) imply that the uncertainty relation not only is minimized on the ground state
$\left|0,1\right\rangle$ but also by all states that are lowest bases of the representations
of the $su(1,1)$ algebra: $\Delta_{l,l+1}=\frac{\hbar^2}{4}$.

\item  The interesting question now is to find the deviation of the minimum
uncertainty for the second bases $\left|l,l+2\right\rangle$
(placed on the second oblique line of Figure 1) of the $l$-representations of $su(1,1)$ algebra.
The expectation values on such states are calculated as
\begin{eqnarray}
&&\hspace{-10mm}\left\langle x\right\rangle_{l,l+2}=\frac{-a_{0}}{2\pi}\left(\Psi(2l+1)-
\frac{1}{2(l+1)}-\ln\left(\frac{-eB_{0}a_{0}^2}{2\pi^2\hbar c}\right)
\right),\nonumber\\
&&\hspace{-10mm}\left\langle x^2\right\rangle_{l,l+2}=\left\langle x\right\rangle_{l,l+2}^2+
\frac{a_{0}^2}{4\pi^2}\left(\frac{l(4l+3)}{2(2l+1)(l+1)^2}+
2(l+1)\zeta(2,2l+1)\right.\nonumber\\
&&\hspace{44mm}-2(2l+1)\zeta(2,2l+2)+(2l+1)\zeta(2,2l+3)\Big{)},\nonumber\\
&&\hspace{-10mm}\left\langle p_{x}\right\rangle_{l,l+2}=i\hbar\frac{2\pi}{a_{0}}(l+1),\hspace{10mm}
\left\langle p_{x}^2\right\rangle_{l,l+2}=\left\langle p_{x}\right\rangle_{l,l+2}^2,\nonumber\\\label{23}
&&\hspace{-10mm}\left\langle xp_{x}\right\rangle_{l,l+2}=-i\hbar(l+1)\left(\Psi(2l+1)
+\frac{l}{2(l+1)^2}-\ln\left(\frac{-eB_{0}a_{0}^2}{2\pi^2\hbar c}\right)\right).
\end{eqnarray}
Therefore, the uncertainty value for these pure states are obtained as
$\Delta_{l,l+2}=\frac{\hbar^2}{4}\left(\frac{3l+2}{l+1}\right)^2$.
Again, we expect this deviation of the minimum uncertainty is
transformed to a known value of the Landau levels in the limit
$l\rightarrow\infty$. We remember that the Landau states in
the symmetric and asymmetric gauges are expressed in terms of Laguerre and
Hermite polynomials, respectively \cite{Mikhailov,Biswas}:
\begin{eqnarray}
&&\hspace{-7mm}\psi_{n,l}^{\mbox{Sy.}}(x,y)
=\sqrt{\frac{\Gamma(n+1)}{\pi\Gamma(n+l+1)}}\left(\frac{x+iy}{\sqrt{2}r_c}\right)^{l}
\frac{e^{-\frac{x^2+y^2}{4r_c^2}}}{\sqrt{2}r_c}L_{n}^{(l)}\left(\frac{x^2+y^2}{2r_c^2}\right),
\hspace{3mm}r_{c}=\sqrt{\frac{\hbar c}{-e B_{0}}}, \nonumber\\\label{24}
&&\hspace{-7mm}\psi_{N}^{\mbox{Asy.}}(x,y)
=\frac{e^{-ik_{y}y}}{2\pi\sqrt{\sqrt{\pi}r_c 2^{N}N!}}e^{-\frac{(x-x_{0})^2}{2r_{c}^2}}
H_{N}\left(\frac{x-x_{0}}{r_{c}}\right),
\hspace{5mm}x_{0}=r_c^2k_{y}-\frac{m E}{eB_{0}^2},
\end{eqnarray}
in which $E$  is the transverse electric field at the surface.
In the symmetric gauge, the rotational symmetry around the
magnetic field direction leads to find the infinite-fold
degeneracy in all the energy levels.
Despite of $n$, $l$ and $N$, $k_{y}$ is a continuous parameter.
The uncertainty value for some of the states of the
symmetric and asymmetric gauges are respectively calculated as
\begin{eqnarray}
&&\hspace{-10mm}\Delta_{0,l}^{\mbox{Sy.}}=\frac{\hbar^2}{4},\hspace{17mm}
\Delta_{1,0}^{\mbox{Sy.}}=\frac{9}{4}\hbar^2,
\hspace{15mm}\Delta_{1,1}^{\mbox{Sy.}}=4\hbar^2,\nonumber\\\label{25}
&&\hspace{-10mm}\Delta_{0}^{\mbox{Asy.}}=\frac{\hbar^2}{4},\hspace{14.5mm}\Delta_{1}^{\mbox{Asy.}}=\frac{9}{4}\hbar^2,
\hspace{12.5mm}\Delta_{2}^{\mbox{Asy.}}=\frac{25}{4}\hbar^2.
\end{eqnarray}
Therefore, the uncertainties $\Delta_{l,l+2}$ for the limit $a_{0}\rightarrow\infty$ ($l\rightarrow\infty$)
are transformed to the uncertainties $\Delta_{1,0}^{\mbox{Sy.}}$
and $\Delta_{1}^{\mbox{Asy.}}$ of the symmetric and asymmetric gauges.

\item  What may be more interesting is the idea that we can obtain the uncertainty values
$\left\{\Delta_{2}^{\mbox{Asy.}}, \Delta_{3}^{\mbox{Asy.}}, \cdots\right\}$
of the Landau levels as limiting cases of
uncertainties for the bases $\left\{\left|l,l+3\right\rangle, \left|l,l+4\right\rangle, \cdots\right\}$
placed on the subsequent oblique lines  of Figure 1. For example, the expectation values on the states
$\left|l,l+3\right\rangle$ are
\begin{eqnarray}
&&\hspace{-7mm}\left\langle x\right\rangle_{l,l+3}=\frac{-a_{0}}{2\pi}\left(\Psi(2l+1)-
\frac{4l+5}{2(l+1)(2l+3)}-\ln\left(\frac{-eB_{0}a_{0}^2}{2\pi^2\hbar c}\right)
\right),\nonumber\\
&&\hspace{-7mm}\left\langle x^2\right\rangle_{l,l+3}=\left\langle x\right\rangle_{l,l+3}^2+
\frac{a_{0}^2}{4\pi^2}\left(\frac{64l^5+344l^4+656l^3+516l^2+125l-13}{4(2l+1)(l+2)(l+1)^2(2l+3)^2}\right.
\nonumber\\
&&\hspace{25mm}+(l+1)(2l+3)\zeta(2,2l+1)-2(2l+1)(2l+3)\zeta(2,2l+2)\nonumber\\
&&\hspace{25mm}+2(2l+1)(3l+4)\zeta(2,2l+3)-2(2l+1)(2l+3)\zeta(2,2l+4)\nonumber\\
&&\hspace{75mm}+(2l+1)(l+2)\zeta(2,2l+5)\Big{)},\nonumber\\
&&\hspace{-7mm}\left\langle p_{x}\right\rangle_{l,l+3}=i\hbar\frac{2\pi}{a_{0}}(l+1),\hspace{10mm}
\left\langle p_{x}^2\right\rangle_{l,l+3}=\left\langle p_{x}\right\rangle_{l,l+3}^2,\nonumber\\\label{26}
&&\hspace{-7mm}\left\langle xp_{x}\right\rangle_{l,l+3}=-i\hbar(l+1)\left(\Psi(2l+1)
+\frac{l(4l+5)}{2(l+1)^2(2l+3)}-\ln\left(\frac{-eB_{0}a_{0}^2}{2\pi^2\hbar c}\right)\right).
\end{eqnarray}
Again we get $\Delta_{l,l+3}=\frac{\hbar^2}{4}\left(\frac{10l^2+19l+8}{(l+1)(2l+3)}\right)^2$. It is obvious that
these uncertainties are
transformed to $\Delta_{2}^{\mbox{Asy.}}=\frac{25}{4}\hbar^2$, in the limit
$l\rightarrow\infty$.
\end{itemize}

\end{document}